# Optical switching of antiferromagnetic domains by nonreciprocal heat current

Takeshi Hayashida,[1,2*] Shunsuke Izumikawa,[3] Kenta Kimura,[3] Carl S. Davies[1,2] and Andrei Kirilyuk[1,2,*]

[1]*HFML-FELIX, Radboud University, Toernooiveld 7, 6525 ED Nijmegen, the Netherlands*
[2] *Institute for Molecules and Materials, Radboud University, Heyendaalseweg 135, 6525 AJ Nijmegen, the Netherlands*
[3]*Department of Materials Science, Osaka Metropolitan University, 599–8531, Osaka, Japan*

* Correspondence authors: takeshi.hayashida@ru.nl; andrei.kirilyuk@ru.nl

## Abstract

What distinguishes front from back? In physics, such directionality emerges only when an underlying symmetry is broken. Antiferromagnets that inherently break both space-inversion and time-reversal symmetries provide a striking example, exhibiting nonreciprocal optical responses that depend on the direction of light propagation. Beyond distinguishing antiferromagnetic domains, we show that this nonreciprocity can deterministically create them. Using mid-infrared light, we demonstrate deterministic switching of antiferromagnetic domains in the magnetoelectric antiferromagnet $LiFePO_4$, where illumination from opposite sides selectively stabilizes opposite domain states. Remarkably, the switching persists over a broad wavelength range rather than being confined to a narrow transition-specific spectral region, overcoming the spectral and material constraints of resonance-based optical switching schemes. The broadband switching originates from the material's intrinsic nonreciprocity through optically generated heat currents. Our results establish nonreciprocity as a general principle for deterministically controlling symmetry-broken phases with light.

## Main

Symmetry determines which physical responses are allowed. Indeed, many functional properties emerge from symmetry breaking[1]. Among the various symmetries relevant to condensed matter systems, space-inversion symmetry ($\mathcal{P}$) and time-reversal symmetry ($\mathcal{T}$) are particularly important. For example, breaking of $\mathcal{P}$ gives rise to ferroelectricity, while breaking of $\mathcal{T}$ leads to ferromagnetism.

What happens when both $\mathcal{P}$ and $\mathcal{T}$ are individually broken, but their combined $\mathcal{PT}$ symmetry remains preserved? This scenario is frequently realized in antiferromagnets, which hereafter we refer to as $\mathcal{PT}$ antiferromagnets. Although antiferromagnets are notorious for being difficult to probe and control because of their vanishing net magnetization, the simultaneous breaking of $\mathcal{P}$ and $\mathcal{T}$ endows them with unique functionalities. A representative example is the linear magnetoelectric (ME) effect[2–4], in which magnetization $M$ (polarization $P$) is induced by an electric field $E$ (magnetic field $H$), described as $M_i = \alpha_{ij}E_j$ ($P_i = \alpha_{ij}H_j$), where $\alpha_{ij}$ is the ME tensor. The corresponding free-energy term, $\Delta F \propto \alpha_{ij}E_iH_j$, allows for domain selection through the simultaneous application of $E$ and $H$.

The ME effect can be even extended to optical frequencies ($\omega$), leading to a variety of rather exotic nonreciprocal optical phenomena, i.e., light propagation direction (**k**) dependent response[5–7]. For example, in materials with off-diagonal ME coupling ($\alpha_{ij}$ with $i \neq j$) at optical frequencies, nonreciprocal directional dichroism (NDD) emerges, in which the transmittance depends on the sign of **k**[8–10], thereby distinguishing front from back. Such optical ME effects facilitate straightforward visualization of antiferromagnetic domains[11–13].

A fundamental question relates to how such nonreciprocal optical responses can be utilized not only for probing but also for controlling antiferromagnetic order. Optical manipulation of antiferromagnets has attracted considerable attention in recent years, and a variety of switching mechanisms have been proposed[14–19]. However, deterministic control of 180° antiferromagnetic domains still remains challenging. Particularly intriguing is the inverse effect of the above-mentioned optical nonreciprocity, in which the same symmetry that governs a nonreciprocal optical response acts as a driving force for deterministic domain switching. This concept has recently been demonstrated at specific lines in the visible and near-infrared spectral ranges[20,21], where the nonreciprocal optical response itself is strongly enhanced by resonant optical transitions. This reliance on specific resonances is not unique to nonreciprocal optical switching but is actually a common feature of many existing approaches for the optical control of ordered states. While they bring advantages such as selectivity and efficiency, resonance-based control also introduces disadvantages, such as material-specificity or demanding requirements for frequency-specificity. It therefore remains unclear whether resonance-enhanced nonreciprocity is truly essential for deterministic switching, or whether nonreciprocity itself can provide a more general route to optical control.

Here, we address this question using intense mid-infrared (IR) pulses in the off-diagonal ME antiferromagnet $LiFePO_4$. We demonstrate deterministic switching of 180° antiferromagnetic domains governed by the light propagation direction, with illumination from opposite sides selectively stabilizing opposite domain states. Strikingly, the effect persists over a broad wavelength range, showing that deterministic switching does not require excitation of a specific optical resonance. We successfully explain the steering of the directional switching in terms of nonreciprocal heat currents, thereby disclosing a general platform for gaining control over order parameters in symmetry-broken antiferromagnetic materials.

$LiFePO_4$ has an olivine-type structure with space group *Pnma*[22]. It is well known as a cathode material of lithium-ion batteries[23], but it also exhibits intriguing magnetic order at low temperatures[24–28]. Below $T_N = 50$ K, $Fe^{2+}$ magnetic moments are antiferromagnetically ordered along the *b* axis, with a slight canting towards the *a* axis[29,30]. Figure 1**a** shows the magnetic structure in which the canting is ignored. The magnetic point group in this case is $mmm'$, which simultaneously break $\mathcal{P}$ and $\mathcal{T}$ symmetries. In this symmetry, two distinct off-diagonal ME tensor components, $\alpha_{ab}$ and $\alpha_{ba}$ ($\alpha_{ab} \neq \alpha_{ba}$), are allowed. There exist two domain states ($L$+ and $L$–) related to each other by either $\mathcal{P}$ or $\mathcal{T}$ operations, and the signs of the ME tensor components are opposite between them. Owing to these off-diagonal ME tensor components, NDD occurs for light propagating along the $c$ axis. It has been reported that NDD reaches its maximum at 1030 nm for linearly polarized light along the $a$ axis, where the difference in the absorption coefficient reaches 25 %[31]. For a fixed propagation direction, the two opposite domain states exhibit different absorption, and the contrast reverses when the propagation direction is inverted. This large difference in absorption enables direct optical imaging of antiferromagnetic domains using a simple experimental configuration (see Figs. 1**b**,**c**).

Using a *c*-cut synthetic bulk single crystal of $LiFePO_4$ (of thickness 157 μm), we performed domain-switching experiments at the free-electron laser facility FELIX in Nijmegen, the Netherlands[32]. As the pump source, mid-IR pulses with wavelengths in the range of 7–19 μm were delivered from FELIX in the form of a single 10 μs-long macropulse consisting of a train of picosecond-long micropulses at a repetition rate of 25 MHz. The macropulse energy at the sample position was typically a few hundreds of micro-Joules. The pump beam was focused onto the sample at normal incidence to a spot with a diameter of approximately 200 μm. Domain imaging was performed using a microscope system based on 300-fs pulses at 1030 nm from an amplified Yb-doped fiber laser, linearly polarized along the *a* axis. Unless otherwise noted, the experiments were performed at 4 K. Further experimental details are provided in the Methods and Extended Data Fig. 1.

Figure 2**a** shows the initial domain structure. To clearly visualize the domain structure, the transmittance image at 4 K ($I_{4K}$) is normalized by that taken in the paramagnetic phase ($I_{paramag}$). The initial state is almost a uniformly "bright" single-domain structure. Although the correspondence between the transmission contrast and the domain states illustrated in Fig. 1**a** cannot be uniquely determined, for simplicity we define in the following the bright (dark) contrast for light propagating parallel to the *c* axis as the *L*+ (*L*–) domain (see the schematic illustration in Fig. 2**a**). The result of irradiation with a single macropulse (14 μm, 300 μJ), linearly polarized along the *b* axis, on the *L*+ domain is shown in Fig. 2**b** (front-side pumping). The irradiated region is switched to the opposite *L*– domain state. The switched domain remains unchanged for at least several minutes after the laser irradiation, confirming that the observed switching can be classified as non-volatile. Moreover, further optical exposures to macropulses have no effect on the switched domain structure.

Next, to perform irradiation from the opposite side, the sample was flipped. Although the domain structure itself remains unchanged by this flipping, the probe beam now propagates antiparallel to the *c* axis, and so the contrast is reversed (Fig. 2**c**). The same region was then irradiated again with the same 14 μm single macropulse (back-side pumping). As a result, the domain now switches back from the *L*– to the *L*+ state (Fig. 2**d**).

These results indicate propagation-direction-dependent domain switching, i.e., *L*+ → *L*– under front-side pumping and *L*– → *L*+ under back-side pumping. We confirmed that no permanent domain change occurs for front-side pumping applied to the *L*– domain or back-side pumping applied to the *L*+ domain. This demonstrates that the domain switching is nonreciprocal with respect to the pump propagation direction.

To further investigate the switching process, and particularly to clarify the conditions that either support or inhibit switching, we performed time-dependent front-side pumping measurements for both the *L*+ and *L*– initial domain states. In the time-dependent measurements performed on the *L*– domain, the pump beam was irradiated onto the same spot for all delay times, since no permanent effects occur in this configuration. In contrast, for measurements on the *L*+ domain, the irradiation position needed to be changed for each delay (see Fig. 3**c** and 3**d** for the initial and final domain structures).

Figure 3**b** shows the result obtained for the *L*– domain at a delay time of 14 μs. Here, the difference between the images before and after irradiation was normalized by the image before irradiation $[(I_{\mathrm{after}} - I_{\mathrm{before}})/I_{\mathrm{before}}]$. After laser irradiation, the intensity of transmitted light increases. The same analysis was performed for each time delay, and the averaged value of $(I_{\mathrm{after}} - I_{\mathrm{before}})/I_{\mathrm{before}}$ within the irradiated region is plotted as blue dots in Fig. 3**e**. The change reaches its maximum at a delay time of 14 ~ 18 μs and subsequently returns toward the initial state in approximately 100 μs.

This transient change of the *L*– domain can be attributed to laser-induced heating of the sample. The green shaded region in Fig. 3**e** represents the expected change in transmission intensity when the sample is heated above $T_{\mathrm{N}}$, indicating that the temperature reaches this regime at delay times of approximately 10–14 μs (see Supplementary Information for details of the estimation of the transmission change expected when the sample is heated above $T_N$ or switched to the opposite domain state).

The red dots in Fig. 3**e** show the results of the measurements for front-side pumping on the *L*+ domain. After laser irradiation, the transmission intensity continues to decrease and reaches, at around 80 μs, the level expected for complete switching to the *L*– domain.

Upon comparing these time-resolved measurements, it is evident that laser-induced heating occurs for both domain states. However, in the case of front-side pumping, switching occurs only in the *L*+ → *L*– direction. On the other hand, the present measurements do not allow us to determine the exact timing of the switching process, i.e., whether the switching occurs at an early stage after irradiation (within ~10 μs) followed by subsequent heating, or instead takes place during the cooling process after the sample has been heated (40–80 μs).

Finally, to gain insight into the possible mechanism of the observed nonreciprocal switching, we measured the wavelength dependence of the switching in the range of 7 – 19 μm. The measurements were performed using front-side pumping on the *L*+ domain. Figure S2 in Supplementary Information summarizes the comparison between the domain images before and after irradiation for different wavelengths. The switching *L*+ → *L*– was observed over a broad wavelength range. It was also confirmed that at wavelengths where *L*+ → *L*– switching was not observed, no permanent switching in the opposite direction (*L*– → *L*+) occurred either.

From the magnitude of the laser-irradiation-induced change, the switching efficiency was evaluated (Fig. 4**b,** see Supplementary Information for details of the evaluation of switching efficiency). Figure 4**a** shows the spectra of the absorption and reflectivity calculated from the room-temperature phonon properties reported in Ref. [33]. Except near 10 μm and 18 μm, where strong absorption and reflectivity peaks originating from transverse optical (TO) phonons are present, switching is observed over a broad wavelength range. A similar trend was observed when the pump polarization was oriented parallel to the *a* axis (Figs. 4**c**,**d**). Importantly, the preferred domain state was determined solely by the light propagation direction and was independent of the pump polarization.

Front-side pumping always induced $L+ \rightarrow L-$ switching, irrespective of whether the pump polarization was parallel to the *a* or *b* axis.

Optical control of ferroic orders in the mid-infrared spectral range has often been associated with resonant phenomena, such as nonlinear phononics driven by infrared-active transverse optical (TO) phonons[34,35] and field enhancement in epsilon-near-zero (ENZ) regions associated with longitudinal optical (LO) phonons[36–38]. In both cases, the response is typically characterized by pronounced resonances at specific wavelengths. In contrast, the switching observed in the present study occurs over a broad wavelength range without exhibiting such sharp resonant behavior. These results suggest that heating, or more generally, the energy deposited into the sample by optical irradiation, plays a dominant role in the switching process. However, it is difficult to conceive how simple heating alone, associated with a scalar rise in temperature, can drive non-reciprocal switching of the vector domain orientation.

To further examine this, we performed simulations of the temperature increase induced by optical irradiation. The black-dashed line in Figs.4**b** and 4**d** shows the wavelength dependence of the calculated temperature increase based on the absorption and reflectivity spectra in Figs. 4**a** and 4**c**, respectively. In this calculation, a top-hat-shaped optical pulse with a duration of 10 μs was assumed, and the volume-averaged temperature within the irradiated region is shown at the time when the temperature reaches its maximum (see Methods for details). The resulting spectrum is in overall agreement with the switching efficiency shown in Figs. 4**b** and 4**d**, reproducing in particular the pronounced reductions in switching efficiency around 10 μm and 18 μm for *b*-polarized pumping and around 9 μm for *a*-polarized pumping. However, for *b*-polarized pumping, the calculated temperature increase does not fully reproduce the switching efficiency at the spectral edge around 7 μm. This discrepancy may be attributed to instability of the FELIX output power near the edges of the accessible wavelength range.

Based on these observations, we discuss possible mechanisms of the switching. As a prerequisite, the experimentally observed **k** dependent nonreciprocal control implies that an effective external field is required whose sign flips with the propagation direction reversal and corresponds one-to-one with the domain state. Such a coupling can be described by a free-energy term $\Delta F \propto \sigma\, \chi_{\mathrm{Ext}}$, where $\sigma$ is a coupling tensor that changes sign between the two domain states and $\chi_{\mathrm{Ext}}$ represents an external field whose sign depends on $\mathbf{k}$. The most straightforward candidate for $\chi_{\mathrm{Ext}}$ is the electromagnetic field of light itself, $\mathbf{E}_\omega \times \mathbf{H}_\omega$. In this case, the coupling term takes the form $\Delta F \propto \alpha_{ab} E_a^\omega H_b^\omega$ (or $\alpha_{ba} E_b^\omega H_a^\omega$) which corresponds to the inverse effect of NDD, as previously reported in the near-infrared spectral range[21]. Since the ME tensor components $\alpha_{ab}$ and $\alpha_{ba}$ change sign between the two domain states, and the relation $\mathbf{k} \parallel \mathbf{E}_\omega \times \mathbf{H}_\omega$ holds, this mechanism satisfies the symmetry requirements.

If the switching is governed by a resonantly-enhanced nonreciprocal optical response, pronounced NDD would be expected in the corresponding wavelength range. However, no such nonreciprocal response has been reported in this spectral region. To directly examine this possibility, we measured the transmittance spectrum in the range of 9 – 14 μm and the temperature dependence of the transmission at 14 μm (see Fig. S3**a** and Fig. S3**b** in the Supplementary Information). We observe no resolvable difference in transmittance between opposite antiferromagnetic domains and no clear anomaly across $T_N$, indicating that any NDD in this spectral range is below the sensitivity of our measurement. This absence of detectable NDD is consistent with the lack of any strong resonant excitations that typically enhance the interaction with $\mathbf{H}_\omega$, such as *d*-*d* transitions[10,39] or electromagnons[40,41].

The absence of detectable NDD suggests that its inverse effect on switching is also expected to be unlikely. Sufficiently intense irradiation combined with moderate absorption of light in the sample may still nominally generate an effective switching force, but it is now more reasonable to consider that the switching is instead primarily mediated by energy-deposition during the optical irradiation. However, the relevant driving force cannot arise from isotropic heating alone, since the experimentally observed switching depends on both the propagation direction of light and the antiferromagnetic domain state. The effective field responsible for switching, $\chi_{\mathrm{Ext}}$, must have the same symmetry as the antiferromagnetic order, namely simultaneous breaking of $\mathcal{P}$ and $\mathcal{T}$ symmetries. A physical quantity generated during optical absorption that satisfies this symmetry requirement is the heat current $\mathbf{J}$, more specifically the heat current along the $c$ axis, $J_c$, which is parallel to the light propagation direction in the present geometry. The corresponding coupling term in the free energy can be written as $\Delta F \propto \tau_c J_c$, where $\tau_c$ is the coupling coefficient that has the same symmetry with $J_c$ and changes sign between the domain states.

Intuitively, the sign of the thermal gradient along the light propagation direction reverses depending on whether the sample is irradiated from the front side or from the back side, and therefore the sign of $J_c$ is also expected to reverse. The heat flow, therefore, acts as a directional symmetry-breaking field, with illumination from opposite sides generating opposite effective fields that select different antiferromagnetic domain states. This behavior is confirmed by a model calculation of the optically induced heat current (see Extended Fig. 2).

If the heat current acts as the switching source, nonreciprocal propagation of heat current is expected. Although such behavior has not previously been identified for $LiFePO_4$, related phenomena have been observed in systems where currents couple to symmetry-broken order parameters. For example, nonreciprocal heat transport has been observed in the multiferroic $TbMnO_3$ where coexisting macroscopic **P** and **M** generate directional responses via $\mathbf{P} \times \mathbf{M}$[42]. Moreover, electric-current-induced switching of antiferromagnetic domains has been reported in metallic antiferromagnets which share the similar symmetry with $LiFePO_4$, where the microscopic mechanism is understood in terms of spin-transfer torque generated by electric current[43–46]. In ferromagnets, temperature gradients are similarly known to generate spin torques through thermally driven spin transport[47–49], and a theoretical study has also suggested that heat-current-induced spin torques can couple to antiferromagnetic order[50]. Incorporating the symmetry of $\mathcal{PT}$ antiferromagnets into such thermal-gradient-induced spin torques may provide a microscopic mechanism for the present switching, although a quantitative theoretical description remains an important subject for future work.

In conclusion, we observed nonreciprocal directional domain switching at mid-IR wavelengths in the ME antiferromagnet $LiFePO_4$. In contrast to previously reported switching in the near-infrared spectral range[21], the present switching does not require a specific resonance and is observed over a broad wavelength range mediated by an optically-generated heat current. This result suggests that **k**-dependent optical switching in this type of system is more universal than previously recognized. The absence of a requirement for a specific optical resonance further suggests that this mechanism may be applicable across a broad range of $\mathcal{PT}$ antiferromagnets.

Many open questions remain for future investigation, including whether nonreciprocal heat transport can be observed in this material, whether domain control can be achieved using other thermal excitation sources, and what the characteristic time scales of the switching dynamics, including possible precession processes, are. This work therefore represents an important milestone toward controlling and understanding antiferromagnetic order using nonequilibrium directional energy flow.

## Methods

### Sample Preparation

Single crystals of $LiFePO_4$ were grown by the flux method[51]. A *c*-cut plate with a thickness of 157 μm was prepared for optical measurements by mechanical grinding and subsequent polishing with diamond slurry on both surfaces.

### Pump-probe imaging using FELIX

Domain-switching experiments were performed at the free-electron laser facility FELIX (Nijmegen, the Netherlands)[32]. A schematic illustration of the experimental setup is shown in Extended Data Fig. 1. Mid-infrared pump pulses with wavelengths between 7 and 19 μm were used. FELIX delivers radiation in the form of a 10-μs-long macropulse consisting of picosecond micropulses at a repetition rate of 25 MHz. The pulse energy at the sample position was typically a few hundred μJ, depending on wavelength.

The pump beam was focused onto the sample at normal incidence to a spot diameter of approximately 200 μm. The pump polarization was set parallel to either the *a* or *b* axis. Front-side and back-side pumping experiments were performed by illuminating the sample from opposite surfaces.

Antiferromagnetic domains of $LiFePO_4$ were imaged using a microscope system based on 300-fs pulses at 1030 nm from an amplified Yb-doped fiber laser. The probe polarization was fixed parallel to the *a* axis. For the time-resolved measurements, the timing of the probe pulse was synchronized to the arrival of the FELIX-delivered macropulse, allowing controlled variation of the pump-probe delay.

## Thermal simulation and heat-current calculation

The temperature evolution following mid-infrared excitation was calculated by numerically solving the heat diffusion equation in an axisymmetric two-dimensional geometry,

$$\rho C_p(T)\frac{\partial T}{\partial t} = \nabla \cdot (\kappa \nabla T) + Q(r, z, t), \tag{1}$$

where $\rho$ is the mass density, $C_p(T)$ is the temperature-dependent heat capacity, $\kappa$ is the thermal conductivity, and $Q(r, z, t)$ represents the absorbed laser power density. The sample was modelled as a cylindrical disk with a thickness of 150 μm and a diameter of 600 μm. The initial temperature was set to 4 K.

The absorbed energy density was calculated from the wavelength-dependent absorption coefficient and reflectivity calculated from the optical spectra in Ref. [[33]]. The spatial profile of the excitation was assumed to follow a Gaussian beam with a diameter of 200 μm and an exponential absorption profile described by the Beer–Lambert law. The temporal profile of the FELIX pulse was approximated by a 10-μs-long top-hat pulse with a total pulse energy of 200 μJ. The heat capacity was taken from experimental $C_p(T)$ data[52], whereas the thermal conductivity was assumed to be isotropic and temperature independent ($\kappa = 5\ \mathrm{W\,m^{-1}K^{-1}}$). Heat exchange with the environment was included through a phenomenological boundary heat-transfer coefficient at the outer radial boundary.

The wavelength dependence of temperature increase shown in Fig. 4**b**,**d** was obtained from the maximum temperature rise averaged over the irradiated volume. The heat-current density was evaluated from the calculated temperature distribution using Fourier's law,

$$\mathbf{J} = -\kappa \nabla T. \tag{2}$$

The resulting heat-current distribution, shown in Extended Data Fig. 2, confirms the emergence of a finite heat-current component along the sample thickness ($J_c$).

## Data availability

The data that support the findings of this study will be publicly available upon publication.

## Competing interests

The authors declare no competing interests.

## Acknowledgements

The images of crystal structures were drawn using the software VESTA[53]. We thank K. Arakawa for fruitful discussion. We thank all technical staff at HFML-FELIX for technical support. We gratefully acknowledge the Nederlandse Organisatie voor Wetenschappelijk Onderzoek (NWO-I) for their financial contribution, including the support of HFML-FELIX. C.S.D. acknowledges support from the European Research Council ERC Grant Agreement No. 101115234 (HandShake), and A.K. acknowledges support from the European Research Council ERC Grant Agreement No. 101141740 (INTERPHON). K.K. acknowledges support from JSPS KAKENHI (Grant No. JP24K00575 and JP26H00677).

## Contributions

T.H. conceived the project, performed the measurements and analyzed the data. C.S.D. and A.K. supervised the project. S.I. and K.K. prepared the sample. T.H. prepared the manuscript in discussions with C.S.D. and A.K. All authors discussed and contributed to the final manuscript.

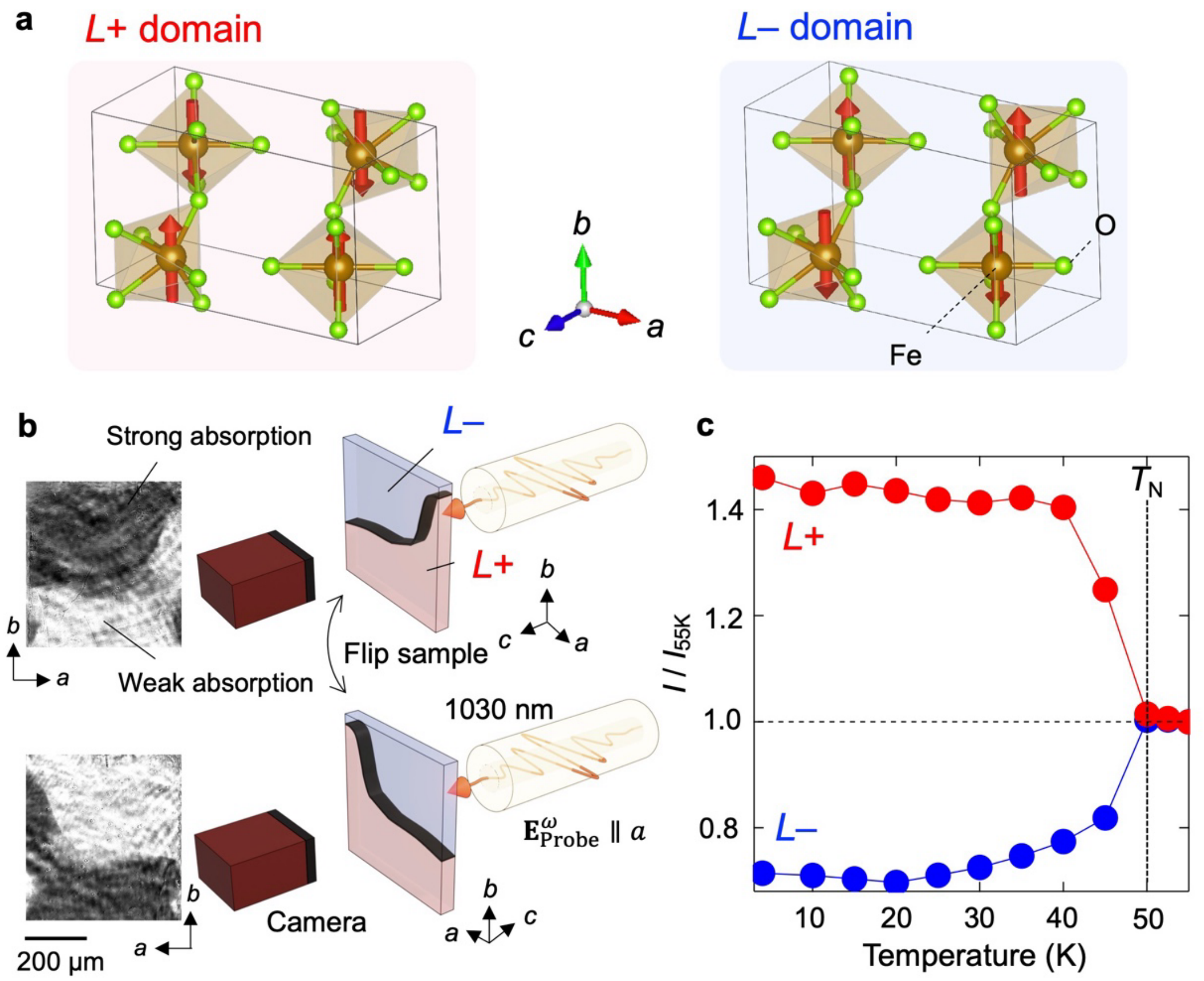

**Fig. 1: Antiferromagnetic domains and nonreciprocal directional dichroism (NDD) in $LiFePO_4$. a**, Magnetic structure of $LiFePO_4$. For clarity, only Fe and O atoms are shown. A collinear antiferromagnetic structure with $Fe^{2+}$ spins (red arrows) oriented along the *b* axis is assumed. The two domain sates, *L*+ (left) and *L*– (right), are related to each other by either $\mathcal{P}$ or $\mathcal{T}$ operations. **b**, Antiferromagnetic domain imaging in $LiFePO_4$ via NDD. The schematic illustrations on the right show the experimental geometry, whereas the images on the left show the spatial distribution of NDD at 4 K obtained using 1030 nm probe light polarized along the *a* axis. To clearly show the domain contrast, the transmittance image at 4 K (below $T_N$) is normalized by that at 55 K (above $T_N$). The bright and dark contrasts are reversed upon flipping the sample, i.e., reversing the sign of **k** (compare top and bottom images). **c**, Temperature dependence of the normalized transmittance for the *L*+ and *L*– domains. For each domain, the transmittance at a given temperature ($I$) is normalized by that at 55 K ($I_{55K}$). The data were obtained in the experimental geometry corresponding to the top panel of **b**.

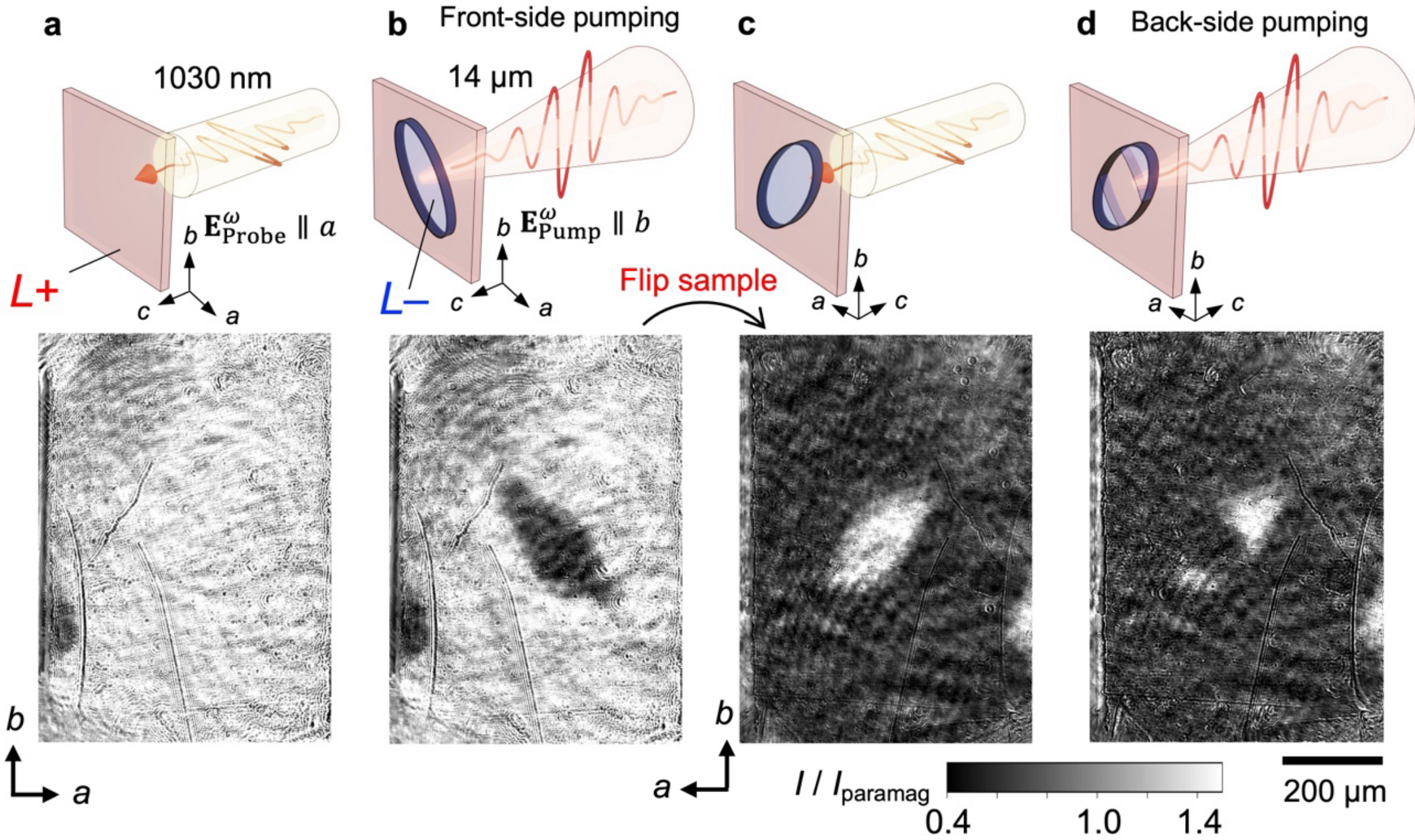

**Fig.2: Propagation-direction-dependent domain switching by mid-IR pulse.** Top panels show the experimental geometry, and bottom panels show domain images obtained by NDD, where the transmittance image at 4 K is normalized by that in the paramagnetic phase. **a**, Initial domain structure, consisting of an almost uniform *L*+ domain. **b**, Domain structure after irradiation of a single 14 μm macropulse polarized along the *b* axis (300 μJ) from the front side (front-side pumping). The irradiated region is switched from the *L*+ to *L*– domain. **c**, Domain image after flipping the sample. The bright and dark contrasts are reversed because of the reversal of **k** of the probe light. **d**, Domain structure after irradiation with the same pulse as in panel **b**, but from the opposite side (back side pumping). The *L*– domain created by the front side pumping was switched back to the *L*+ domain.

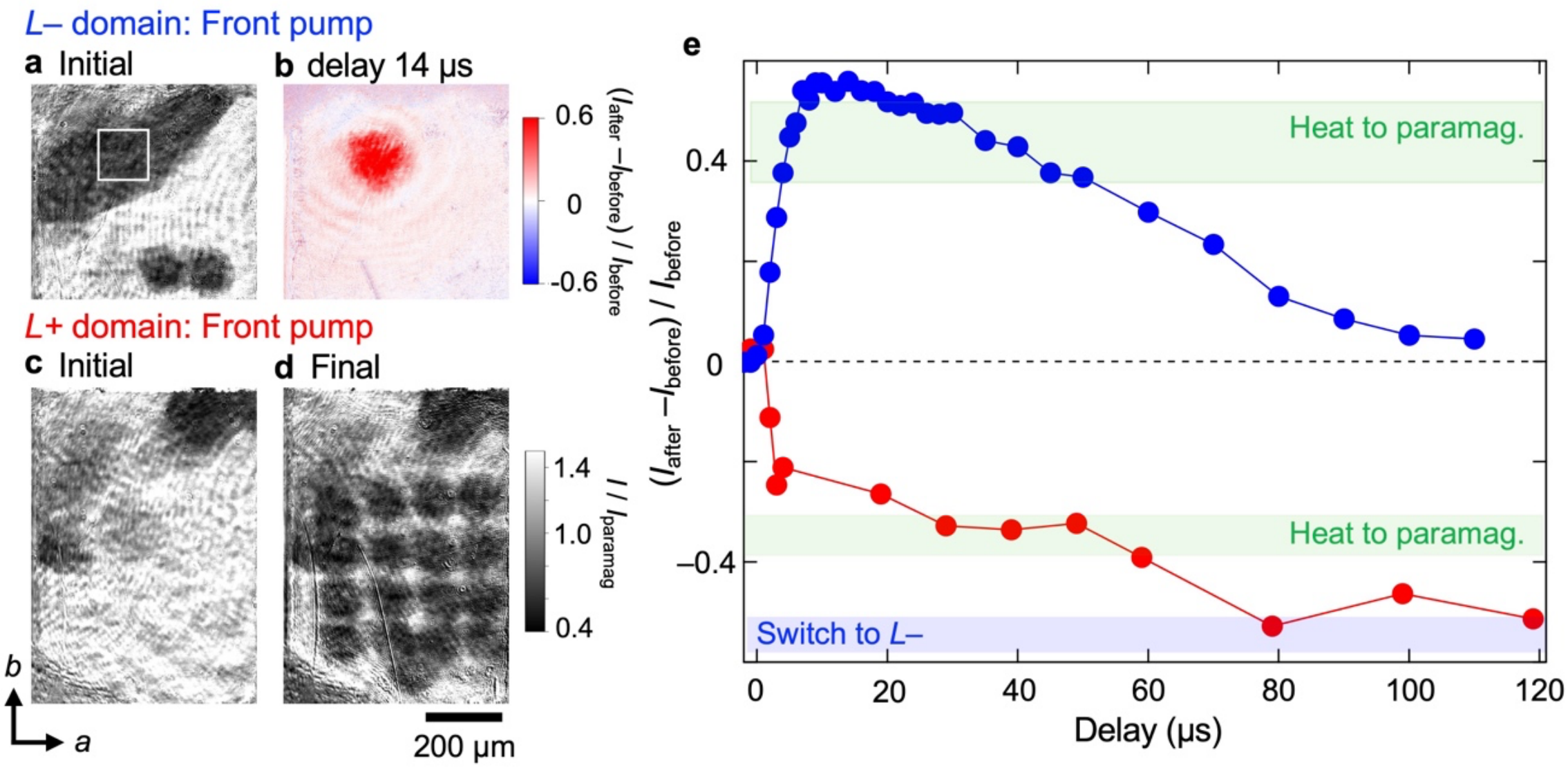

**Fig. 3**: **Time-resolved evolution of pump-induced transmission changes. a**, Initial domain structure for front-side pumping of the *L*– domain. A single 14 μm macropulse (300 μJ) was irradiated onto the region indicated by the white box. **b**, Pump-induced transmission change at a delay time of 14 μs for the *L*– domain. The image is calculated as $(I_{\mathrm{after}} - I_{\mathrm{before}})/I_{\mathrm{before}}$, where $I_{\mathrm{after}}$ and $I_{\mathrm{before}}$ are the domain images acquired before and after the pump irradiation, respectively. **c**,**d**, Initial (**c**) and final (**d**) domain structures for front-side pumping of the *L*+ domain. The pump pulse was applied to a different location for each delay time. **e**, Time dependence of the pump-induced transmission change. Red and blue dots show the results for the *L*+ and *L*– domains, respectively. The plotted values correspond to the spatial average of $(I_{\mathrm{after}} - I_{\mathrm{before}})/I_{\mathrm{before}}$ within the irradiated region. For the *L*– domain, the measurements were repeated five times at each delay time and averaged. The green shaded region indicates the expected transmission change associated with heating from the antiferromagnetic phase to the paramagnetic phase. The blue shaded region indicates the expected transmission change for complete switching from the *L*+ to the *L*– domain.

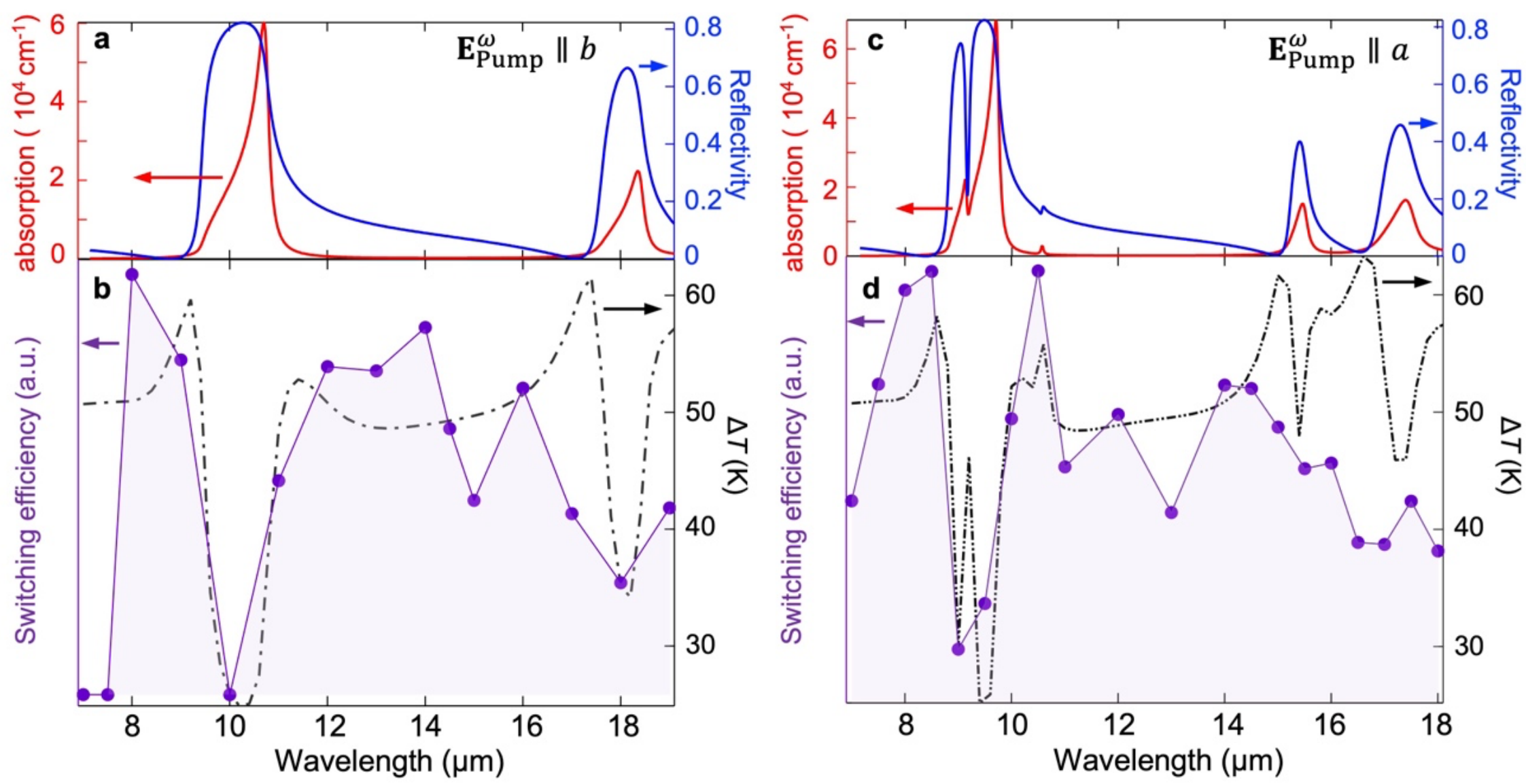

**Fig.4**: **Spectral dependence of the switching efficiency. a**,**c**, Absorption (red) and reflectivity (blue) spectra for light polarized along the *b* (**a**) and *a* (**c**) axes. The spectra were calculated from room-temperature phonon parameters obtained by spectroscopic ellipsometry in Ref. [33]. **b**,**d**, Wavelength dependence of the switching efficiency for pump polarization along the *b* (**b**) and *a* (**d**) axes. The black dashed line shows the estimated temperature increase induced by the pump pulse, calculated using the absorption and reflectivity spectra shown in **a** and **c**.

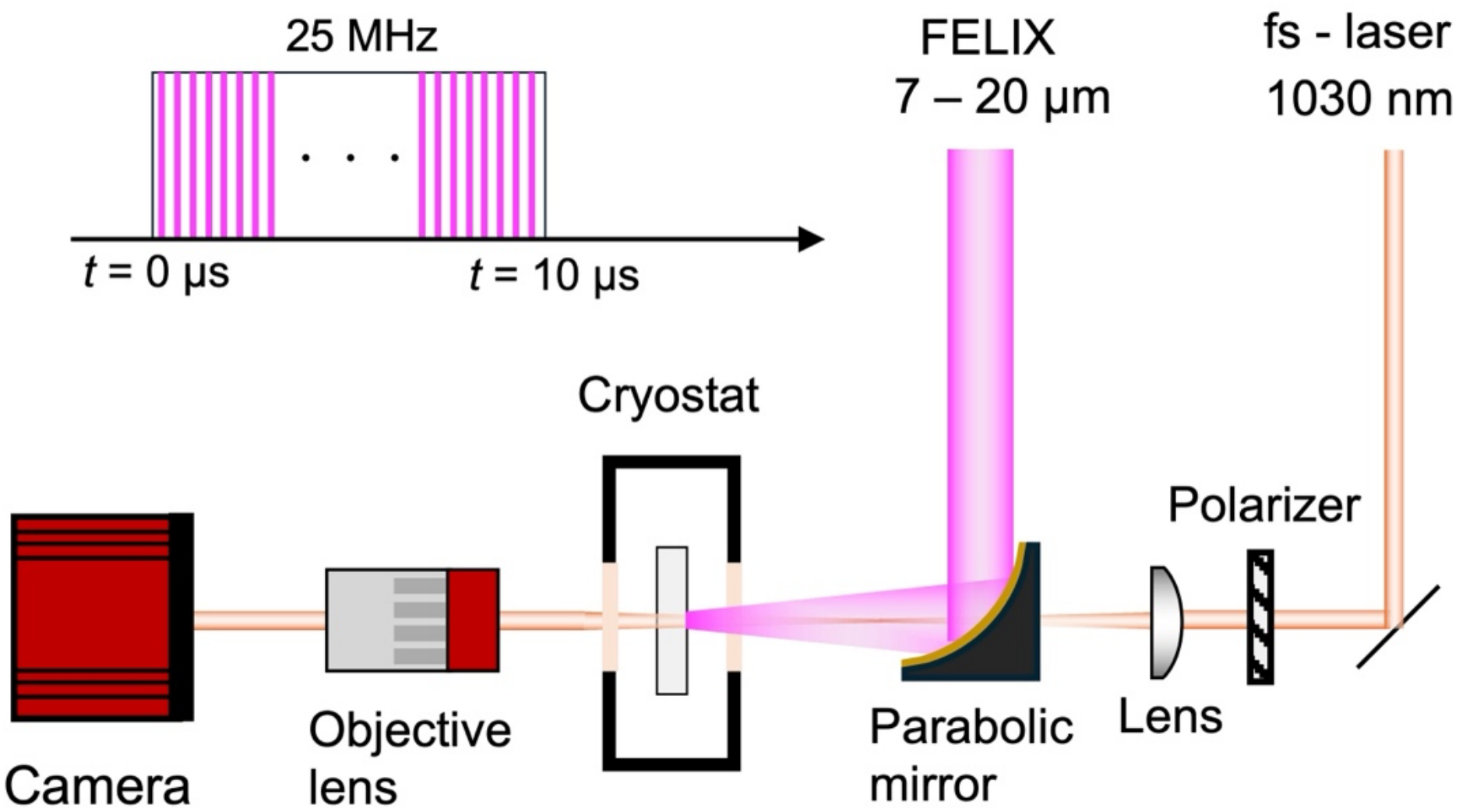

**Extended Data Fig. 1: Schematic illustration of the experimental setup.** Mid-infrared pump pulses from FELIX were used to excite the sample, while antiferromagnetic domains were imaged using 1030 nm probe pulses.

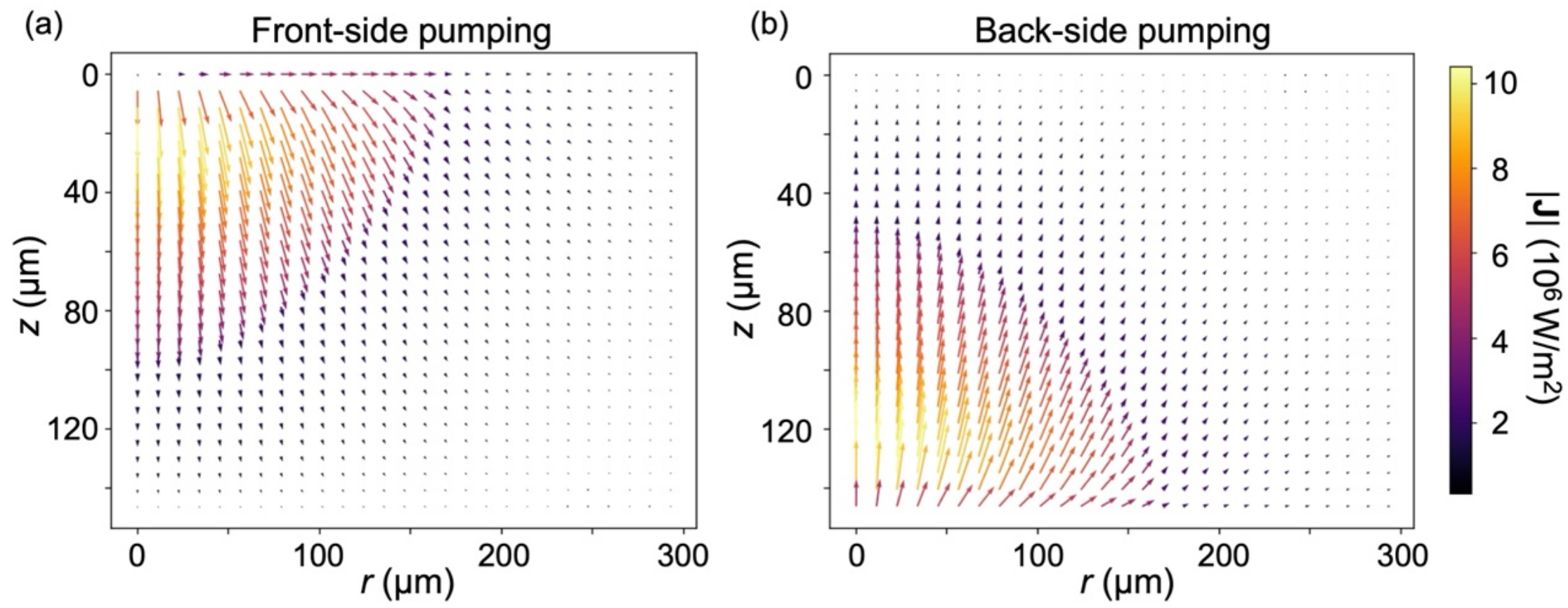

**Extended Data Fig. 2: Simulated heat-current distribution by the laser irradiation**. Simulated heat-current distribution at $t$ = 10 μs after excitation by a 14 μm pulse polarized along the $b$ axis, assuming a pulse energy of 200 μJ: (a) front-side pumping and (b) back-side pumping. The horizontal axis represents the radial coordinate $r$, while the vertical axis represents the depth $z$ within the sample, with $z$ = 0 corresponding to the front-side sample surface. Arrows indicate the direction of the heat current density **J**. A pronounced heat-current along the thickness direction is generated near the center of the excitation spot, and its direction reverses between front-side and back-side pumping.

Supplementary Information for

# Optical switching of antiferromagnetic domains by nonreciprocal heat current

Takeshi Hayashida,[1,2*] Shunsuke Izumikawa,[3] Kenta Kimura,[3] Carl S. Davies[1,2] and Andrei Kirilyuk[1,2,*]

*[1]HFML-FELIX, Radboud University, Toernooiveld 7, 6525 ED Nijmegen, the Netherlands*
*[2] Institute for Molecules and Materials, Radboud University, Heyendaalseweg 135, 6525 AJ Nijmegen, the Netherlands*
*[3]Department of Materials Science, Osaka Metropolitan University, 599–8531, Osaka, Japan*

* Correspondence authors: takeshi.hayashida@ru.nl; andrei.kirilyuk@ru.nl

This file includes:
Supplementary Note 1-3
Figure S1-S3

**Supplementary Note 1 | Estimation of transmission changes induced by heating and domain switching**

We estimate the transmittance changes associated with two processes relevant to the pump-induced dynamics discussed in the main text: (i) heating of the *L*+ and *L*– domains to the paramagnetic phase and (ii) switching of the *L*+ domain to the *L*– domain. Figure S1 shows the normalized transmission image obtained by dividing the transmission image at 4 K by that at 55 K. The *L*+ and *L*– domains appear as bright and dark regions, respectively. Transmission values were sampled at multiple positions within each domain. From these measurements, we obtained average normalized transmission values of $I_{4\mathrm{K},L+}/I_{55\mathrm{K}} = 1.54 \pm 0.09$ and $I_{4\mathrm{K},L-}/I_{55\mathrm{K}} = 0.70 \pm 0.04$ where the errors represent the standard deviations of the sampled values. Using these values, we calculated the expected changes in transmission associated with heating to the paramagnetic phase and with switching from the *L*+ domain to the *L*– domain, which are shown as the shaded regions in Fig. 3**e** of the main text. The finite widths of the shaded regions reflect the standard deviations given above.

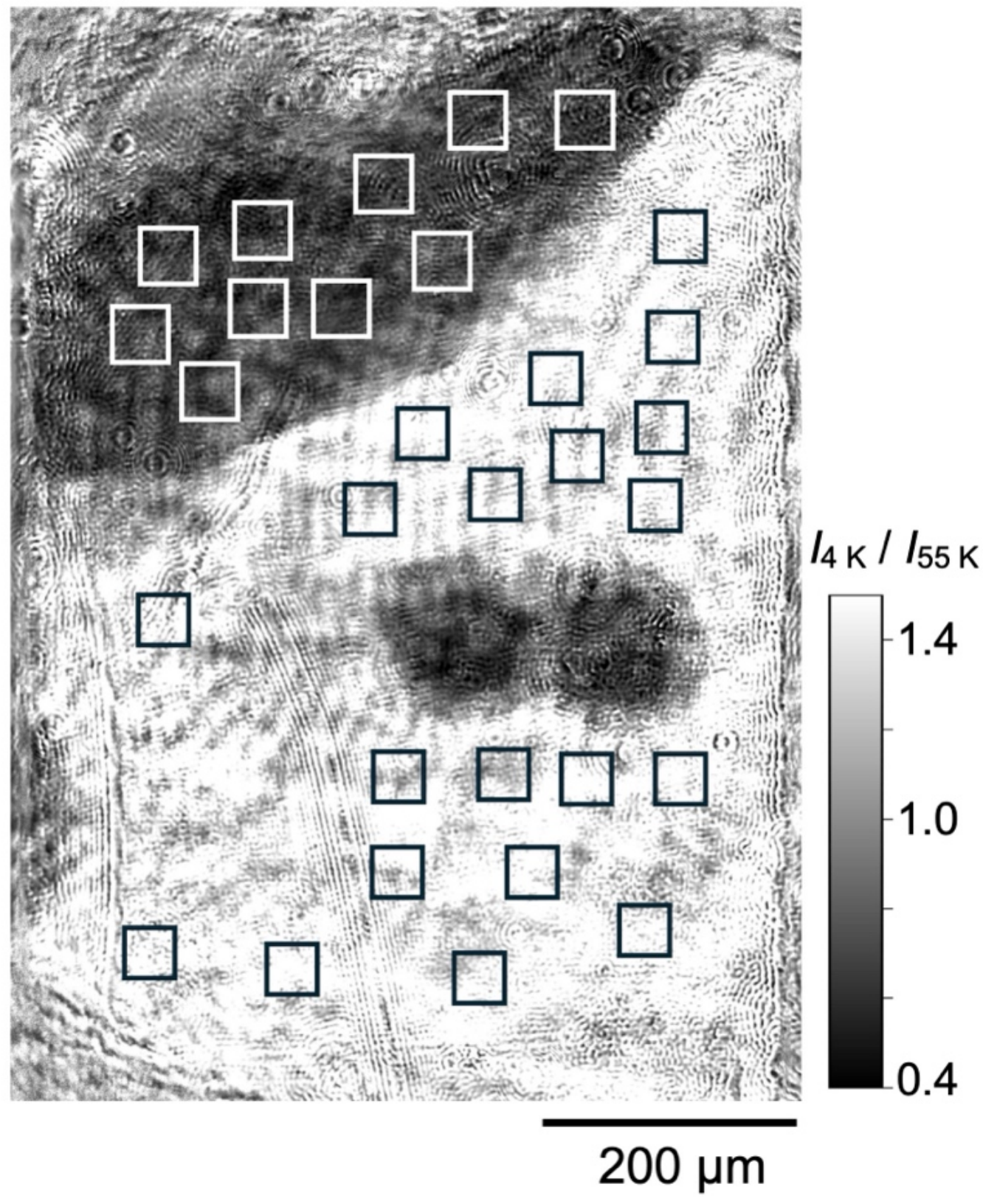

**Fig. S1: Normalized transmission image obtained by dividing the transmission image at 4 K by that at 55 K.** Black and white boxes indicate the regions used for statistical analysis of the *L*+ and *L*– domains, respectively.

**Supplementary Note 2 | Evaluation of switching efficiency for different pump wavelengths**

The wavelength dependence of the switching was investigated by irradiating the *L*+ domain with single FELIX macropulses while varying the pump wavelength in the front-side pumping. Measurements were performed for pump polarizations parallel to both the *a* and *b* axes. Representative results are shown in Figs. S2**b**–**e** and **f**–**i**. For each wavelength, the change in the transmission image induced by the pump pulse was evaluated as $(I_{\mathrm{after}} - I_{\mathrm{before}})/I_{\mathrm{before}}$, where $I_{\mathrm{before}}$ and $I_{\mathrm{after}}$ are the transmission images acquired before and after the pump irradiation, respectively. Since the images were recorded several seconds after the excitation, the resulting contrast reflects non-volatile changes of the antiferromagnetic domains. To quantify the switching efficiency, $(I_{\mathrm{after}} - I_{\mathrm{before}})/I_{\mathrm{before}}$ was averaged over the regions indicated by the black boxes in Figs. S2**b**–**e** and **f**–**i**. The resulting values were further normalized by the pump pulse energy shown in Figs. S2**a**,**f**. These normalized values were used as the switching efficiency shown in Fig. 4**b,d** of the main text. It should be noted that the pulse energy used for the normalization was obtained from the FELIX diagnostic station and therefore does not correspond to the absolute pulse energy at the sample position.

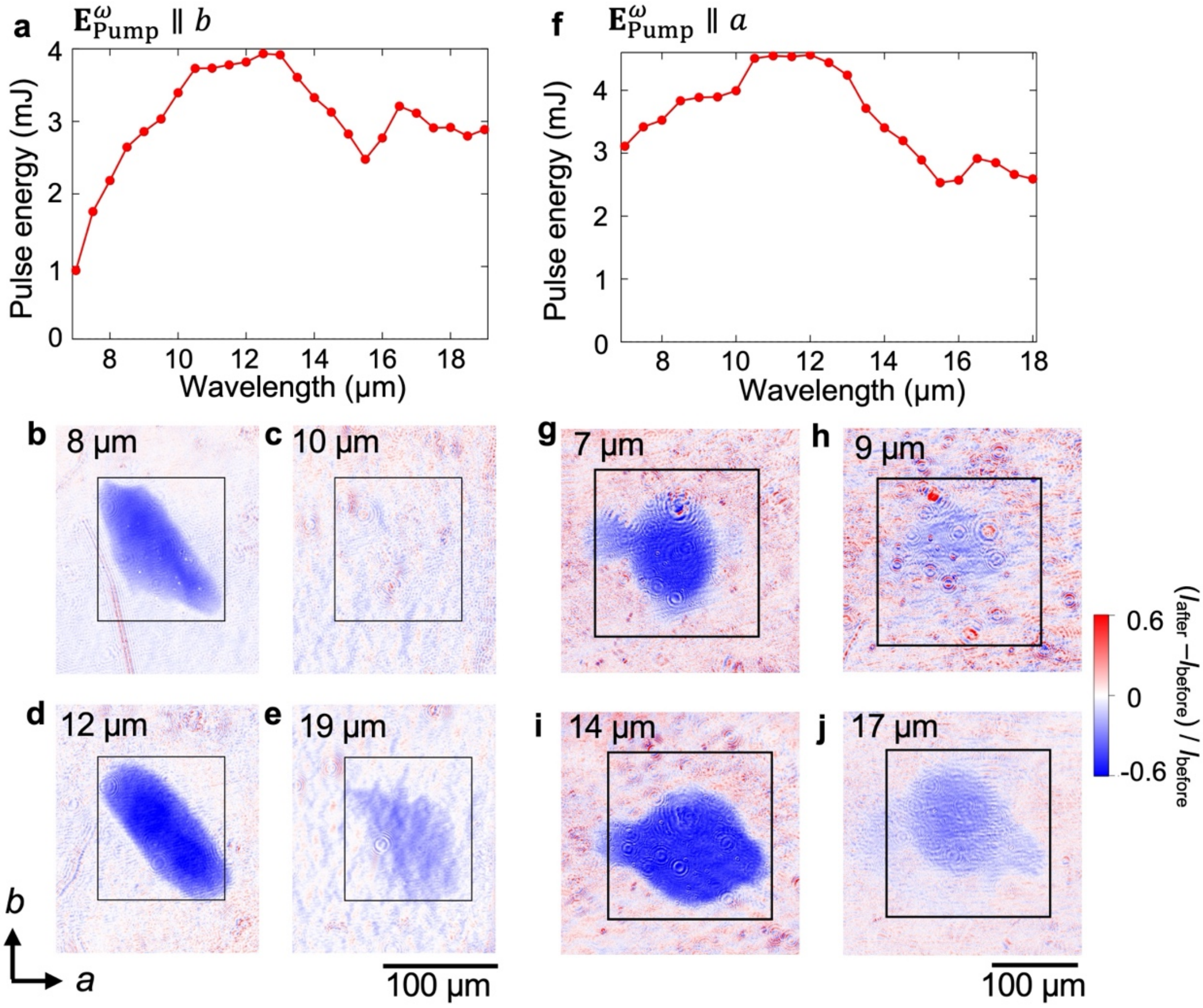

**Fig. S2: Evaluation of switching efficiency for different pump wavelengths**. **a**,**f** Pump pulse energy measured at the FELIX diagnostic station for the $b$-polarized pump (**a**) and $a$-polarized pump (**f**). **b**–**e**, Representative images showing the non-volatile transmission change induced by a single FELIX macropulse with the pump polarization parallel to the $b$ axis at wavelengths of 8 μm (**b**), 10 μm (**c**), 12 μm (**d**), and 19 μm (**e**). The images show $(I_{\text{after}} - I_{\text{before}})/I_{\text{before}}$, where $I_{\text{before}}$ and $I_{\text{after}}$ are the transmission images acquired before and several seconds after the pump irradiation, respectively. The black boxes indicate the regions used for evaluating the switching efficiency. **g**–**j**, Representative domain images obtained with the pump polarization parallel to the $a$ axis at wavelengths of 7 μm (**g**), 9 μm (**h**), 14 μm (**i**), and 17 μm (**j**), displayed in the same manner as in **b**–**e**.

**Supplementary Note 3 | Mid-infrared transmission measurements of *L*+ and *L*– domains**

To examine whether nonreciprocal directional dichroism (NDD) is present within the wavelength range where domain switching was observed, transmission spectra were measured using FELIX in the wavelength range of 9–14 μm with the light polarized along the *b* axis. The measurement geometry was identical to that used in the front-side pumping experiments.

To compare the transmission of the *L*+ and *L*– domains at the same position, the following procedure was employed. First, the transmission spectrum of an *L*+ domain was measured by illuminating the sample with a focused FELIX beam at a pulse energy sufficiently below the switching threshold (below 10 μJ). The transmitted intensity was recorded using a photodiode while the wavelength was varied. Subsequently, the same location was irradiated with a 14 μm pulse of 300 μJ, which switched the domain from *L*+ to *L*–. The transmission spectrum was then measured again under the same low-energy conditions. The resulting spectra are shown in Fig. S3**a**. No clear difference in transmission between the *L*+ and *L*– domains was observed in the wavelength range from 11 to 14 μm, where the deterministic switching was observed.

To further investigate possible optical signatures associated with the antiferromagnetic order, the temperature dependence of the transmission at 14 μm was measured on an *L*+ domain while heating the sample through $T_N$. The result is shown in Fig. S3**b**. No distinct anomaly was observed at $T_N$.

These results indicate that, within the present experimental sensitivity, no detectable nonreciprocal directional dichroism was observed in the wavelength range where domain switching occurs.

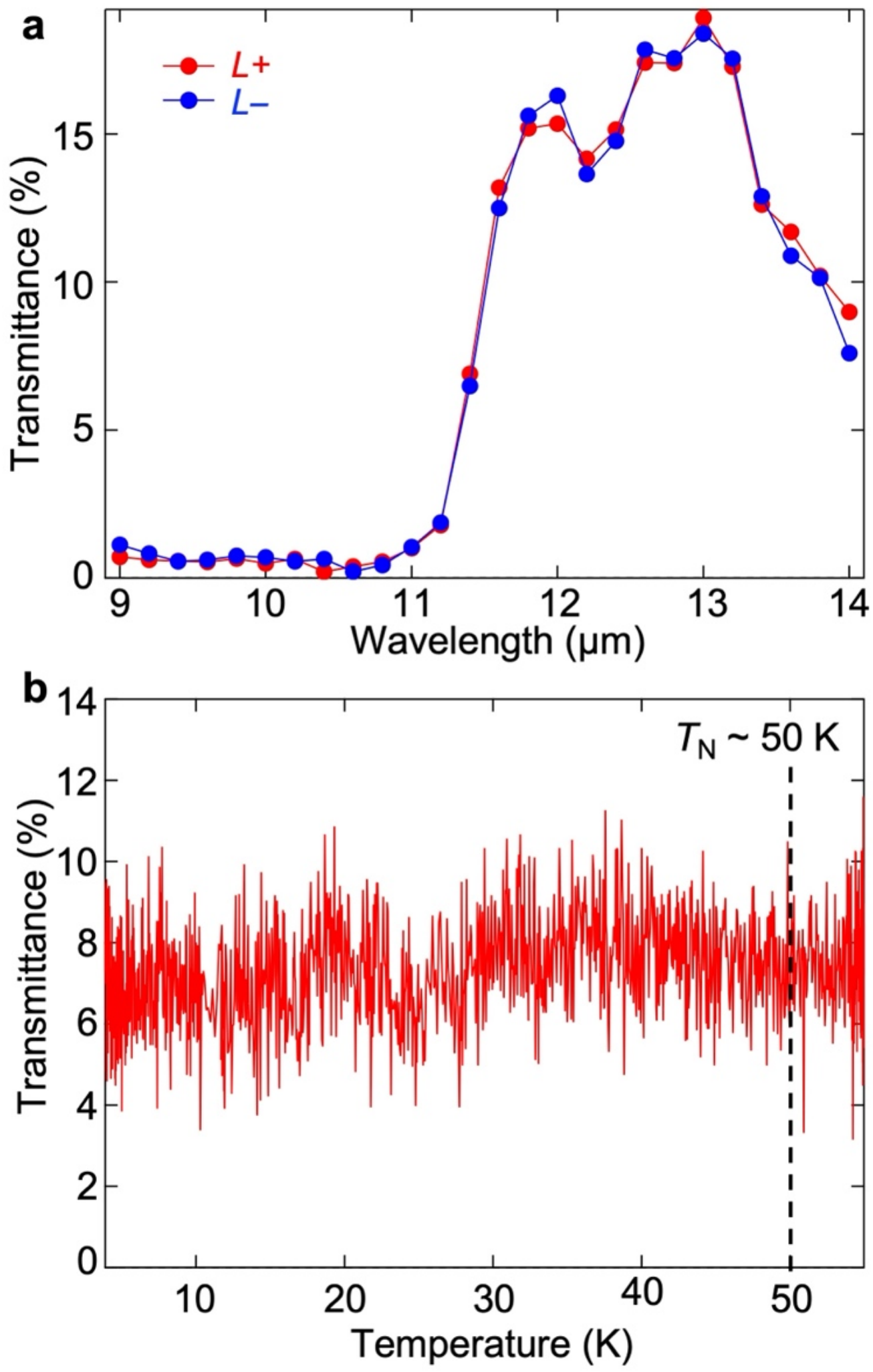

**Fig. S3: Mid-infrared transmission measurements of the *L*+ and *L*– domains. a**, Transmission spectra of the *L*+ and *L*– domains measured using FELIX with the light polarized along the *b* axis. No clear difference in transmission is observed between the two domains. **b,** Temperature dependence of the transmission at 14 µm measured on the *L*+ domain. No distinct anomaly is observed at $T_N \sim 50$ K.